\documentclass[aps,twocolumn,showpacs,superscriptaddress,groupedaddress]{revtex4-1} 
\usepackage[charter]{mathdesign}
\usepackage{amsmath,bm}
\usepackage{graphicx}	
\usepackage{xcolor}

\newcommand\Sigmav{\mathbf{\Sigma}}
\newcommand\dg{\dagger}
\newcommand\dgf{{\phantom\dagger}}
\newcommand\Gammav{\mathbf{\Gamma}}
\newcommand\thetav{\bm{\theta}}
\newcommand\epsv{\bm{\varepsilon}}
\newcommand\tv{\mathbf{t}}
\newcommand\Gv{\mathbf{G}}
\newcommand\kvt{\mathbf{\tilde k}}

\begin{document}

\title{Spontaneous nematicity triggered by impurities in a Hubbard model for the cuprates}

\author{Alexandre Foley}
\affiliation{D\'epartment de physique and RQMP, Universit\'e de Sherbrooke, Sherbrooke, Qu\'ebec, Canada J1K 2R1}

\author{David S\'en\'echal} 
\affiliation{D\'epartment de physique and RQMP, Universit\'e de Sherbrooke, Sherbrooke, Qu\'ebec, Canada J1K 2R1}

\date{\today}

\begin{abstract}
The physical properties of high-$T_c$ superconductors are affected by spatial inhomogeneities introduced by impurities. In addition, superconductivity and electronic nematicity seem intertwined in these materials. To address these questions, we apply inhomogeneous cluster dynamical mean field theory to the study of superconductivity in the Hubbard model, in the presence of a repeated impurity, at zero temperature. We find that the superconducting phase is shifted slightly away from half-filling due to the presence of the impurity. This can be explained by a competition with the Mott insulating phase which then persists at finite doping. In addition, the impurity triggers the appearance of spontaneous nematicity. The nematic order parameter follows a dome shape as a function of doping, similar to that of the superconducting order parameter, and increases with impurity potential.
\end{abstract}

\pacs{}
\maketitle


High-$T_c$ superconducting copper oxides are by and large defect rich, and at the same time remarkably resilient to disorder~\cite{Alloul:2009fr}.
Mean-field analyses predict that $d$-wave superconductivity is very sensitive to the presence of disorder and would be completely suppressed at relatively low impurity concentrations~\cite{fehrenbacher_gap_1994,tolpygo_universal_1996}.
It has been surmised that this resilience to disorder is an effect of electron correlations, as for instance indicated by theoretical studies based on the $t-J$ model~\cite{Garg:2008yq,Chakraborty:2014rt,Tang:2015fk}.
A dynamical mean-field theory study of the disordered Hubbard model points to enhanced impurity screening as a part of the explanation~\cite{Tanaskovic:2003fj}.

In this paper we apply an inhomogeneous implementation of cluster dynamical mean field theory (I-CDMFT~\cite{Charlebois:2015qy}) to the problem of a periodically repeated impurity in the square lattice Hubbard model, in order to see its effect on the superconducting phases at zero temperature. We find that the superconducting dome, which extends all the way to half-filling in pure systems, at least for small clusters~\cite{Senechal:2005,Kancharla:2008vn}, is shifted towards higher doping by the presence of impurities. 
We stress that $d$-wave superconductivity appears dynamically in our approach and is not put in by hand.

In addition, we observe spontaneous nematicity in the superconducting phase due to the presence of the impurity. This phenomenon can be seen both in the profile of the site charge $n_i=\langle c^\dg_{i\sigma} c_{i\sigma}\rangle$ and in that of the bond charge
$B_{ij}=\langle c^\dg_{i\sigma} c_{j\sigma} + \mbox{H.c.}\rangle$.
Nematicity, or broken $C_4$ symmetry, has been observed in the pseudogap phase of cuprate superconductors~\cite{Hinkov:2008uq,Daou:2010kx,Cyr-Choiniere:2015fk} as well as within the superconducting phase itself by scanning tunneling microscopy~\cite{Kohsaka:2007ly,Lawler:2010vn,Fujita:2014zr}, sometimes in conjunction with a bond density-wave.
Whereas it is not clear to what extent impurities have an essential role in breaking $C_4$ symmetry in the experiments, we find that spontaneous nematicity disappears as the strength of the impurity potential goes to zero. 


\paragraph*{Model and Methodology}

The one-band Hubbard model with site impurities is defined by the following Hamiltonian:
\begin{equation}\label{eq:ImpHmod}
H = \sum_{i,j,\sigma} t_{ij} c^\dagger_{i\sigma}c^\dgf_{j\sigma}
+ U \sum_{i} n_{i\uparrow}n_{i\downarrow}
+ \sum_{i} h_i n_i
\end{equation}
where $t_{ij}$ is the hopping amplitude between sites $i$ and $j$, $U$ is the on-site Coulomb repulsion and $h_i$ is a site-dependent impurity potential.
The chemical potential $\mu$ is included as the diagonal of the matrix $t_{ij}$.
We will adopt a band structure adequate for YBCO~\cite{andersen_lda_1995}: The nearest-neighbor hopping amplitude will be set to $t=1$ (this defines the energy scale); the second-neighbor (diagonal) amplitude is $t'=-0.3$ and the third-neighbor amplitude is $t''=0.2$.
We will adopt the intermediate coupling value $U=8$ for the interaction.

\begin{figure}[b]
\begin{center}
\includegraphics[scale =0.9]{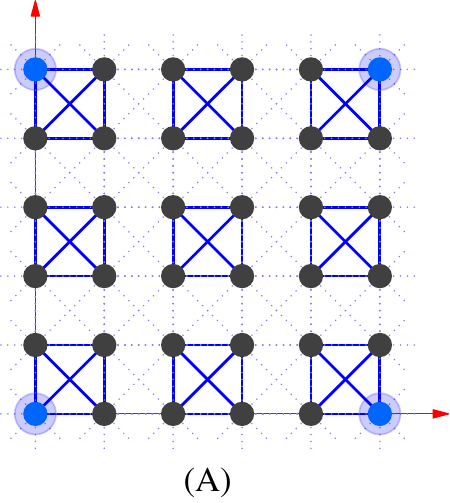}\hfil
\includegraphics[scale =0.9]{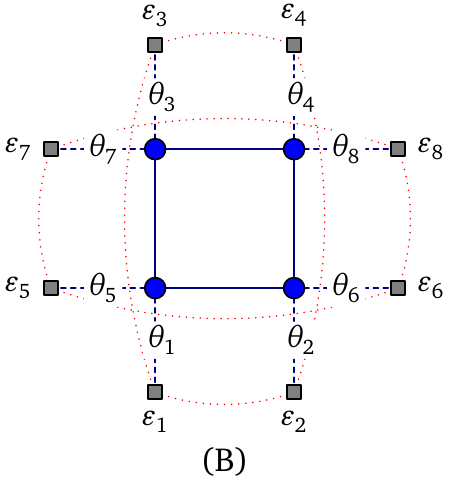}
\caption{(Color online) (A) A super unit cell with 9 clusters. The impurities are located at the corners. Because of periodicity, this amounts to four adjacent sites, but located on different clusters.
(B) The CDMFT bath configuration of a single 4-site cluster. Bath sites are pictured as gray squares. Some bath parameters may be related by symmetry, depending on the cluster's location within the super unit cell. Pairing operators within the bath are shown as red dotted lines, without symbols attached.}\label{fig:system}
\end{center}
\end{figure}

The impurity configuration will consist of a repeated pattern of four neighboring sites with a non-zero impurity potential $h_i=h$, all other sites having $h_i=0$.
The repeated unit (or super unit cell) is illustrated on Fig.~\ref{fig:system}A and has 36 sites.
The four impurity sites could have been placed in the center, but instead have been placed at the four corners. 

We use cluster dynamical mean field theory~\cite{Lichtenstein:2000vn,Kotliar:2001,Senechal:2012kx} (CDMFT) with an exact diagonalization solver at zero temperature.
The basic assumption behind this method is that the electron self-energy is well approximated by that of a small cluster (here four sites) immersed in a bath of $N_b$ non-interacting orbitals whose parameters are determined self-consistently in order to best represent the environment of that cluster.
On each four-site cluster we thus define an Anderson impurity model with Hamiltonian
\begin{multline}
H_{\rm AIM} = \sum_\alpha t_{c,\alpha\beta} c^\dagger_\alpha c^\dgf_\beta
+ U \sum_{i} n_{i\uparrow}n_{i\downarrow} \\
+ \sum_{\alpha,r}\left(\theta_{\alpha r}c^\dg_\alpha a^\dgf_r + \mathrm{H.c.}\right)  
+ \sum_{r,s} \varepsilon_{rs} a_r^\dg a^\dgf_s 
\end{multline}
where Greek indices are a composite of site and spin: $\alpha=(i,\sigma)$.
The one-body matrix $\tv_c$ contains hopping terms restricted to the cluster, the chemical potential, and the impurity potential $h_i$.
The bath orbitals (spin included) are labelled by latin indices $r,s$ and the correponding annihilation operators are $a_r$.
The Green function $\Gv_c$ of the cluster itself, in matrix form, can be shown to have the form
\begin{equation}\label{eq:hybrid3}
\Gv_c^{-1}(\omega) = \omega - \tv_c - \Gammav(\omega)-\Sigmav(\omega)
\end{equation}
where the matrix $\Gammav(\omega)=\thetav(\omega-\epsv)^{-1}\thetav^\dg$, the so-called hybridization function, is expressed in terms of the hybridization matrix $\thetav$ between cluster and bath, and of the one-body terms $\epsv$ within the bath itself.

In order to represent the superconducting state within that formalism, we work in Nambu space: $c_\alpha = (c^\dgf_{i\uparrow},c^\dg_{i\downarrow})$ and $a_r = (a^\dgf_{j\uparrow},a^\dg_{j\downarrow})$. The Green function matrix $\Gv_c$ has a block structure, with the off-diagonal block constituting the Gork'ov function $\mathbf{F}_c(\omega)$. Superconductivity is seeded in the system by introducing pairing between the bath sites, i.e., off-diagonal elements of opposite spins within the block matrix $\epsv$.
For a given set of bath parameters, the cluster Green function $\Gv_c(\omega)$, and consequently the associated self-energy $\Sigmav(\omega)$, are computed via the Lanczos method.

\begin{figure}[t]
\begin{center}
\includegraphics[scale=0.9]{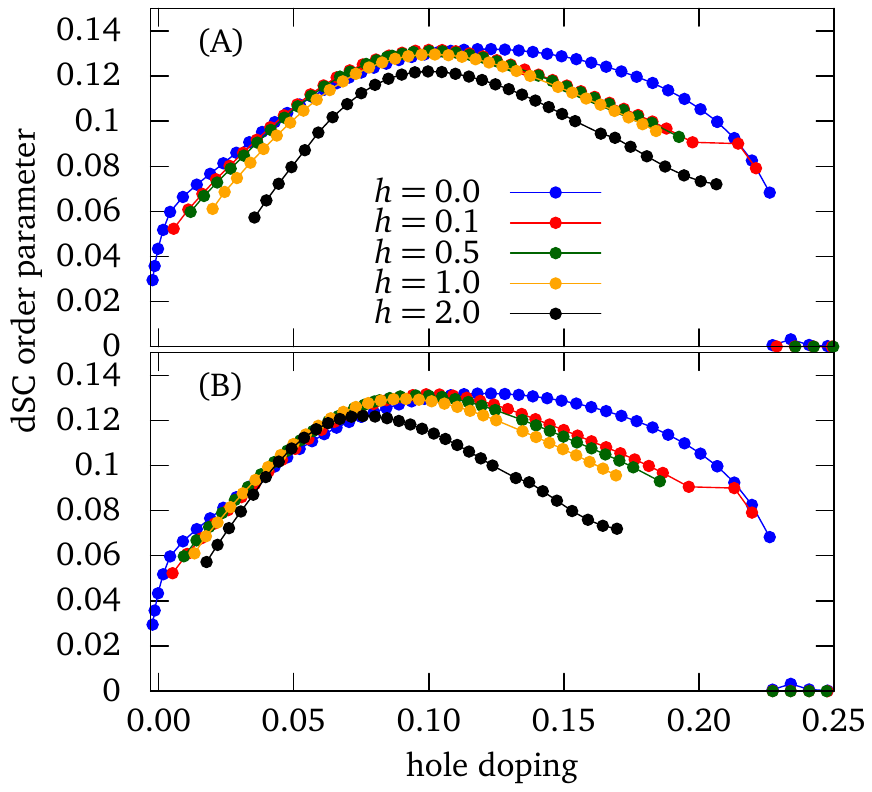}
\caption{(Color online) Top panel: The $d$-wave order parameter as a function of hole doping, for different impurity strengths $h$. Bottom panel: The same, except that the hole doping is computed without the impurity sites.}\label{fig:Dsc_ybco_eximp_pd}
\end{center}
\end{figure}

Since we are dealing with an inhomogeneous system, or at least with a system with a longer periodicity, we need to use several distinct clusters that together form a super unit cell, here of size $N=36$. This technique was used for studying the effect of correlations in the problem of a non-magnetic impurity in graphene~\cite{Charlebois:2015qy}.
Each of the nine clusters of Fig.~\ref{fig:system}A has the structure shown on Fig.~\ref{fig:system}B. 
The four sites are hybridized with 8 bath orbitals, each with a bath energy $\varepsilon_i$ ($i=1,\dots,8$) (the diagonal elements of the matrix $\epsv$, the same for up and down spins except for a sign). The hybridization amplitudes are denoted $\theta_i$. In addition, we add 4 pairing terms between the bath orbitals, represented by red dotted lines on Fig.~\ref{fig:system}B. 
This amount to 20 bath parameters per cluster. Most of these parameters are constrained by the reflexion symmetry of the super unit cell. Vertical and horizontal mirror symmetry is used to reduce the number of free parameters to 60; this constraint still leaves room for a spontaneous breaking of the $C_4$ symmetry, including $d$-wave superconductivity or electronic nematicity.

Let us say a few words about the self-consistency condition that determines the bath parameters.
The super unit cell defines a superlattice and degrees of freedom may be labelled by an index $\alpha$ within the super unit cell and a reduced wavevector $\kvt$ within the Brillouin zone of that superlattice, $N$ times smaller than the original Brillouin zone. Tensors like the Green function or the self-energy then take the form of wavevector-dependent $2N\times 2N$ matrices $\Gv_c(\omega)$ and $\Sigmav(\omega)$ (if we include Nambu indices).
The self-energy $\Sigmav$ is simply the direct sum of the self-energies of the separate clusters: It is block diagonal. The infinite system's self-energy is approximated by that direct sum, and its Green function then takes the following form:
\begin{equation}
\Gv(\kvt,\omega) = \left[ \omega - \tv(\kvt) - \Sigmav(\omega) \right]^{-1}
\end{equation}
where $\tv(\kvt)$ is a mixed (super unit cell -- wavevector) representation of the one-body matrix of the infinite system (including chemical and impurity potentials).
One can define a projection of this infinite-system Green function onto the super unit cell, via a partial Fourier transform:
\begin{equation}
\bar\Gv(\omega) = \int_\kvt\left[ \omega - \tv(\kvt) - \Sigmav(\omega) \right]^{-1}
\end{equation}
The self-consistency condition demands that this projected Green function coincide with that obtained from the exact diagonalization procedure: $\bar\Gv(\omega)=\Gv_c(\omega)$.
This condition cannot be satisfied exactly, because we only have a finite number of bath parameters at our disposal (60, as mentioned above); in practice, a merit function is optimized (see Ref.~\cite{Charlebois:2015qy} for details).

\paragraph*{Results}

In Fig.~\ref{fig:Dsc_ybco_eximp_pd}A, we show the $d$-wave order parameter computed from I-CDMFT for different impurity strengths $h$. The solutions were obtained without probing antiferromagnetism.
The most striking feature is the abrupt change in the overdoped region when going from $h=0$ (pure system) to even a very small impurity strength.
On the other hand, the change is very gradual in the underdoped region.
As the system is weakly correlated in the overdoped region, it is tempting to conclude that electronic correlations make the system more resilient to disorder.
On the underdoped side, we note that the superconducting phase seems simply shifted towards larger doping as $h$ increases, without much effect on the maximum value, except when $h=2.0$. In Fig.~\ref{fig:Dsc_ybco_eximp_pd}B, we show the same data, with a different definition of average density (or doping) which excludes the four sites directly affected by the impurity. As the different curves are rather similar for low values of $h$ in the underdoped region, it is as if the role of the impurity, at first approximation, was to exclude volume from the system.
(In the remainder of this paper, we will use the usual definition of average density $\langle n\rangle$ or hole doping $x=1-\langle n\rangle$, taking into account all sites.) 

\begin{figure}
\begin{center}
\includegraphics[width=\hsize]{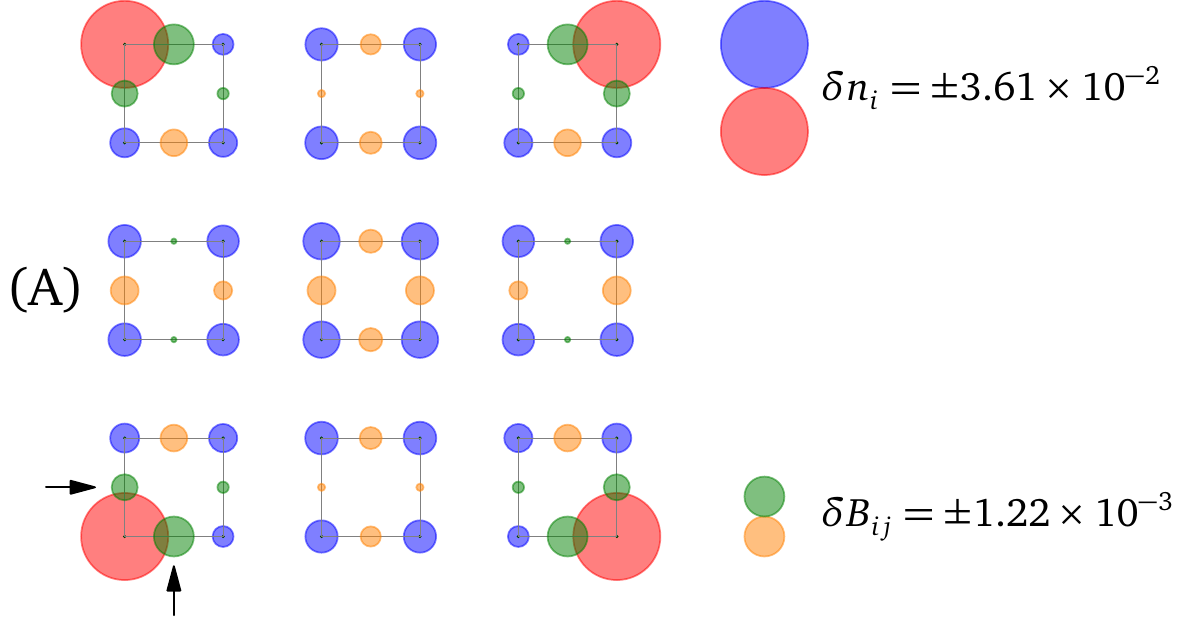}
\vglue5mm
\includegraphics[width=\hsize]{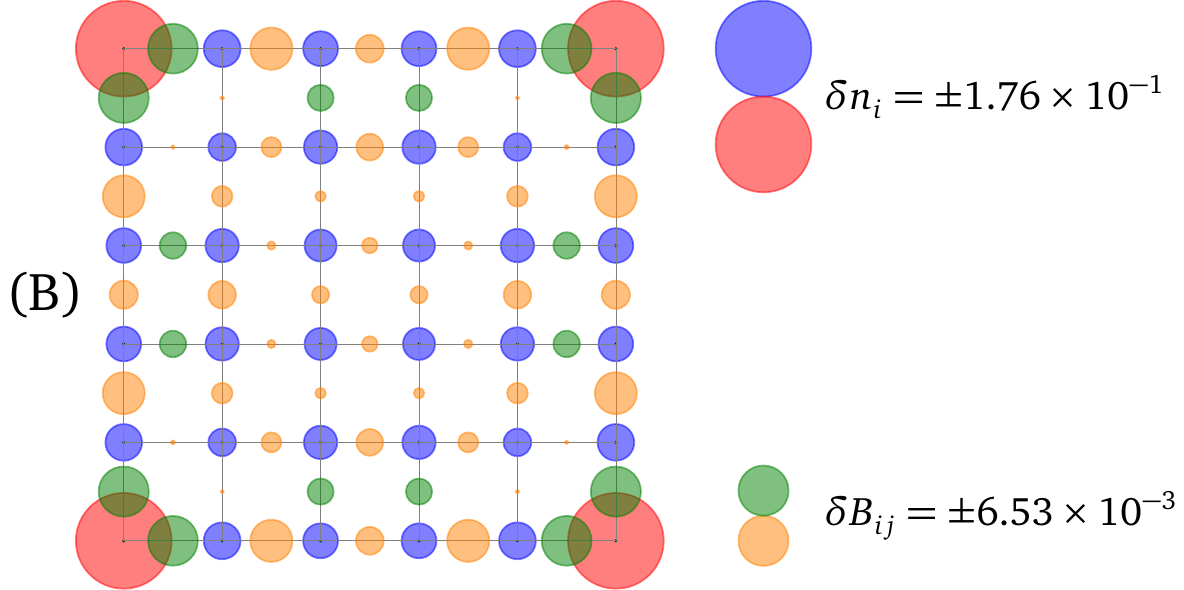}
\end{center}
\caption{(Color online) Local site and bond charges within the super unit cell. The impurity potential is located on the four corners. The area of the red and blue circles located at the sites is proportional to the deviation $\delta n$ of the site charge from its average value (red means a deficit and blue an excess). The yellow and green circles are the deviations from its average value of the bond charge $B_{ij}=\langle c_i^\dagger c_j + \mathrm{H.c.}\rangle$ on each link. The scales are indicated on the side. Panel (A) shows I-CDMFT data (only in-cluster links are shown) and Panel (B) mean-field data from a non-interacting system (see text for details). The average density is $\langle n\rangle = 0.92$, the impurity potential is $h=0.5$ and the superconducting $d$-wave order parameter is $\Psi = 0.14$ in panel (A) and $\Psi = 0.10$ in panel (B).}\label{fig:profile_SC}
\end{figure}

In the absence of impurity, the SC phase extends all the way to half-filling, where the order parameter goes to zero when $U$ is larger than a critical value~\cite{Kancharla:2008vn}. 
The effect of the impurity is to shift the SC dome towards higher doping.
This shift can be understood through the effect of the impurity on the Mott insulator: In the absence of an impurity potential, and if broken symmetry phases are suppressed, any finite amount of doping will make the system metallic. But in the presence of impurities that are attractive to the charge carriers (holes), like in the case we considered, the system will remain an insulator until some critical doping is reached. This is supported by Fig.~\ref{fig:Dsc_ybco_eximp_pd}B, where we can observe that the superconducting dome appears as soon as the charge carriers are numerous enough to leak out of the impurity sites.

In Fig.~\ref{fig:profile_SC}, we show local observables in the SC phase within the super unit cell at $8\%$ hole doping (almost optimally doped in these CDMFT simulations).
In panel (A), we show the data obtained by I-CDMFT for $U=8$. 
On panel (B) we show data for a quadratic, mean-field Hamiltonian, with the same band parameters and impurity strength as the interacting system, but where a superconducting mean-field is imposed.
The values of this mean field and of the chemical potential are chosen in such a way that the observables are close to those of the strongly correlated system.
The local observables are computed from the Green function $\mathbf{G}$ obtained from either approach (CDMFT or mean-field):
\begin{equation}
\langle O\rangle = i\int_{-\infty}^\infty\frac{d\omega}{2\pi}\int_\kvt \mathrm{Tr}\left[\mathbf{O}(\kvt) \mathbf{G}(\kvt,i\omega)\right]
\end{equation}
where $O$ is any local observable one-body matrix and $\mathbf{O}$ the corresponding $2N\times 2N$ matrix. The frequency integral is taken along the imaginary axis, with suitable subtraction of the leading singularity as $\omega\to\infty$.

For all practical purpose, the superconducting order parameter is homogeneous in both the mean-field and strongly correlated solutions and is not shown. On the other hand, the modulation of the site and bond charges are different in the two solutions. 
On Fig.~\ref{fig:profile_SC}B the site density deviation $\delta n_i$ displays a Friedel-like oscillation with a maximum density on the first neighbors of the impurity sites. 
In the strongly correlated system (Fig.~\ref{fig:profile_SC}A) we observe the opposite: the maximum density is located on the sites furthest from the impurity.
A simple physical explanation lies in the effective superexchange interaction between neighboring sites within the Hubbard model: This interaction lowers the ground state energy, and the system will seek to concentrate the density at sites where this interaction is most effective, i.e., not next to the impurity sites, where the density is depressed by the impurity potential.

\begin{figure}[tbh]
\begin{center}
\includegraphics[scale=0.9]{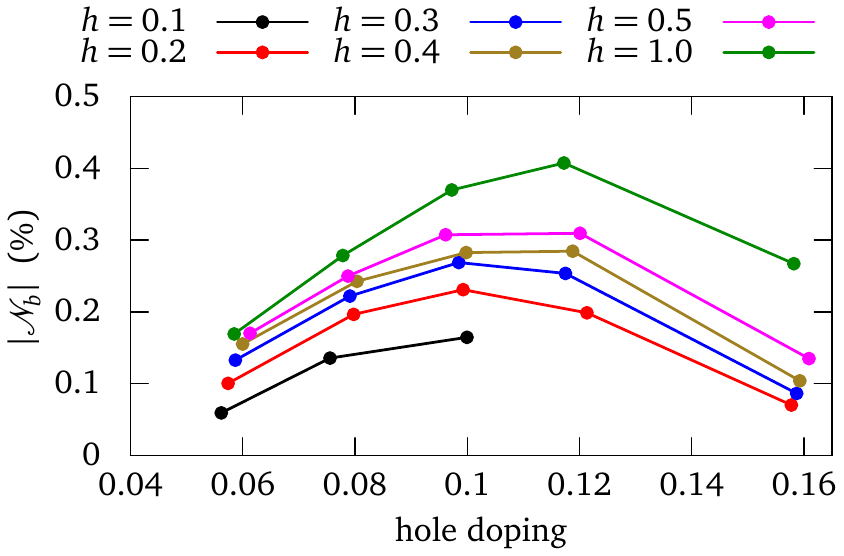}
\caption{(Color online) Nematicity order parameter $|\mathcal{N}_b|$ as a function of doping for various values of the impurity strength, within the superconducting solutions. The order parameter is defined as the difference of the local charges on the $x$ and $y$ bonds next to an impurity site.}
\label{fig:nema}
\end{center}
\end{figure}

Perhaps the most interesting feature of the solutions we found is the breakdown of rotational symmetry: the system has spontaneous nematicity. This can be seen both from the site and bond charge densities in Fig.~\ref{fig:profile_SC}A: The figure is not equivalent to its rotation by 90 degrees. Note that nematicity should not be confused here with an effect of Friedel oscillations. For instance, the anisotropy in bond charges displayed along the bottom-center plaquette on Fig.~\ref{fig:profile_SC}A is not a signature of nematicity, but the difference between this plaquette and a rotated version of the left-center plaquette is.
This asymmetry is not observed in the mean-field profile of Fig.~\ref{fig:profile_SC}B.

We define a nematicity order parameter as follows: 
If $B_x$ is the local charge $B_{ij}$ on the lower left $x$-bond of Fig.~\ref{fig:profile_SC}A and $B_y$ the same in the $y$ direction (indicated by black arrows), then the bond nematicity order parameter is defined as
\begin{equation}
\mathcal{N}_b = \frac{B_x-B_y}{B_x+B_y}
\end{equation}
We show on Fig.~\ref{fig:nema} the value of $|\mathcal{N}_b|$ as a function of hole doping for various impurity strengths; the sign of the nematic order parameter is spontaneous, and is not correlated to that of the SC order parameter: Only the absolute value is shown.
Note that this nematicity, invisible in the pure state, increases with disorder strength. 
It is triggered by the presence of impurities, and was not found in the pure system.
The nematic order parameter follows a dome shape similar to that of the SC order parameter and increases nonlinearly with the impurity strength.
This is to be contrasted with the results of Ref.~\cite{fang_local_2013}, obtained in the variational cluster approximation (VCA), where spontaneous nematicity is found in the SC phase in a pure system and extends very far from half-filling.
We could likewise define a nematic order parameter based on site (not bond) charges, for instance by considering the densities $n_x$ and $n_y$ on the two nearest neighbors of the impurity site at the lower-left corner of Fig.~\ref{fig:profile_SC}A. The site nematic order parameter $\mathcal{N}_s = (n_x-n_y)/(n_x+n_y)$ shows a similar shape to the bond nematic order parameter $\mathcal{N}_b$ defined above, except that its amplitude is roughly one third the size.

\paragraph*{Conclusion}

We have applied I-CDMFT to superconductivity in the Hubbard model with a repeated impurity. The resulting phase diagram shows that the superconducting dome is shifted towards higher doping by the presence of the impurity. This can be explained by the survival of the Mott insulating phase at finite doping in these circumstances.
Impurities also cause the presence of electronic nematicity.
The nematic order parameter follows a dome shape similar to that of the SC order parameter and increases with
impurity strength.

\begin{acknowledgments}

Discussions with A.-M.S. Tremblay and Simon Verret are gratefully acknowledged. 
Computing resources were provided by Compute Canada and Calcul Qu\'ebec.
This research is supported by NSERC grant no RGPIN-2015-05598 (Canada).

\end{acknowledgments}


%

\end{document}